\documentclass[preprint,authoryear,12pt]{elsarticle}

\usepackage{epsfig}
\usepackage{rotating}

\usepackage{amssymb}
\usepackage{graphicx, amsmath}
\usepackage{amsthm,amssymb}
\usepackage{multirow}
\usepackage{longtable}
\usepackage{cite}
\usepackage{amsfonts}

\usepackage[ps2pdf,%
a4paper=true,%
breaklinks=true,%
colorlinks=true,%
pdfauthor={First Author et al.},%
pdftitle={Template for manuscripts in Advances in Space Research}%
]{hyperref}

\journal{Advances in Space Research}

\begin{document}

\begin{frontmatter}



\title{Medium Resolution Near-Infrared Spectra of the Host Galaxies of Nearby Quasars}


\author[khu]{Huynh Anh Nguyen Le}\ead{huynhanh7@khu.ac.kr}
\author[khu]{Soojong Pak\corref{cor}}\ead{soojong@khu.ac.kr}
\author[snu]{Myungshin Im}
\author[kasi]{Minjin Kim}
\author[khu]{Chae Kyung Sim}
\author[iaa,iaab]{Luis C. Ho}

\address[khu]{School of Space Research, Kyung Hee University\\
 1732 Deogyeong-daero, Giheung-gu, Yongin-si, Gyeonggi-do, 446-701, Korea}
\address[snu]{Department of Physics and Astronomy, Seoul National University\\
 Center for the Exploration of the Origin of the Universe (CEOU), Seoul, Republic of Korea}
\address[kasi]{Korea Astronomy and Space Science Institute, Daejeon, Republic of Korea}
\address[iaa]{Kavli Institute for Astronomy and Astrophysics, Peking University, Beijing 100871, China}
\address[iaab]{Department of Astronomy, Peking University, Beijing 100871, China}

\cortext[cor]{Corresponding author}

\begin{abstract}

We present medium resolution near-infrared host galaxy spectra of low redshift  quasars, PG~$0844+349$ (z=0.064), PG~$1226+023$ (z=0.158), and PG~$1426+015$ (z=0.086). The observations were done by using the Infrared Camera and Spectrograph (IRCS) at the Subaru 8.2 m telescope. The full width at half maximum of the point spread function was about 0.3 arcsec by operations of an adaptive optics system, which can effectively resolve the quasar spectra from the host galaxy spectra. We spent up to several hours per target and developed data reduction methods to reduce the systematic noises of the telluric emissions and absorptions. From the obtained spectra, we identified absorption features of Mg~I (1.503 $\mu$m), Si~I (1.589 $\mu$m) and CO (6-3) (1.619 $\mu$m), and measured the velocity dispersions of PG~$0844+349$ to be 132 $\pm$ 110 km~s$^{-1}$ and PG~$1426+015$ to be 264 $\pm$ 215 km~s$^{-1}$. By using an $M_{BH}$$-$$\sigma$ relation of elliptical galaxies, we derived the black hole (BH) mass of PG~$0844+349$, $log(M_{BH}/M_{\odot})$  = 7.7 $\pm$ 5.5 and PG~$1426+015$, $log(M_{BH}/M_{\odot})$ = 9.0 $\pm$ 7.5. These values are consistent with the BH mass values from broad emission lines with an assumption of a virial factor of 5.5.

\end{abstract}


\end{frontmatter}

\parindent=0.5 cm

\section{Introduction}

Nearby galaxies have bulge with supermassive black holes \citep{Richstone98}. Understanding the link between the supermassive black holes and their host galaxies is important in studying the formation and evolution of the galaxies. The relation of $M_{BH}$$-$$\sigma$ has been discovered, in which $M_{BH}$ is the mass of supermassive black hole and $\sigma$  is the stellar velocity dispersion of the bugle (e.g., \citealp{Ferrarese01}; \citealp{Gebhardt00a}; \citealp{Gebhardt00b}).

Nevertheless, the measurements of stellar velocity dispersion of host galaxy are difficult in optical bands because of the presence of young stars in the host galaxy. Absorption lines in optical bands such as Mg~b at 517 nm and Ca triplet at 850 nm are diluted by continuum. Therefore, it is necessary to use stellar lines in other wavebands in measuring velocity dispersion. CO bandheads in near-infrared (NIR) have been suggested to be the best in studying the velocity dispersion of nearby galaxies \citep{McConnell11}. In addition, NIR stellar lines have the potential of explaining for the relation between supermassive back holes and their host galaxies.

In this paper, we present the medium resolution host galaxy spectra of nearby quasars in H-band obtained at the Subaru telescope. Thanks to the advantages of using adaptive optics technology, we can isolate the quasar spectra from the host galaxy spectra. The obtained spectra with medium resolution can be used to determine the stellar velocity dispersions in the bulge of the host galaxies, and to estimate the supermassive black hole masses.

Section 2 of this paper shows the observation processes. The detailed data reduction processes of NIR quasar spectra are presented in section 3. Results and discussions are shown in section 4. Section 5 is the conclusion.

\section{Observations}

The observations were performed at the Subaru 8.2 m telescope using the IRCS \citep{Kobayashi00} operated with the Adaptive Optics (AO), AO36 \citep{Hayano08}, on 2003 February 11 and 2004 April 3 and 4. The average AO-assisted point spread function was 0.3 arcsec.

\subsection{Observation of Quasars}

We observed three nearby quasars, PG~$0844+349$, PG~$1226+023$, and PG~$1426+015$. Table \ref{observation} shows the log of the observations. In 2003, we observed PG~$0844+349$ only. The slit width was 0.3 arcsec with R = $10^{4}$, and the position angle of the slit was 0 deg. The echelle setting of the spectrograph was in $H+$ setting ($1.47-1.82$ $\mu$m), and the total integration time was about two hours with each exposure of 180 sec. The observations were done in an {\it Nod-off-slit} mode. We first observed the target and then moved the telescope to the nearby background sky. The sequences of the observations were $object$ $\rightarrow$ $sky$ $\rightarrow$ $sky$ $\rightarrow$ $object$.

In 2004, we observed three targets: PG~$0844+349$, PG~$1226+023$, and PG~$1426+015$. The slit width was 0.6 arcsec with R = 5 $\times$ $10^{3}$. The echelle settings of the spectrograph were in $H-$ setting ($1.46-1.83$ $\mu$m) and $H+$ setting, and the total integration time was one hour for each target. Other instrument settings and the observation modes were the same as in 2003.

\subsection{Standard Stars and Template Stars}

We observed A0 V type standard stars to correct the telluric absorption lines in the target spectra. In addition, bright template stars (H $<$ 5 mag) in spectral classes of G, K, and M with the luminosity class of III are used to measure the velocity dispersions of the host galaxies.

\section{Data Reduction}

\begin{figure*}[thp]
\centering
\includegraphics[width=45ex]{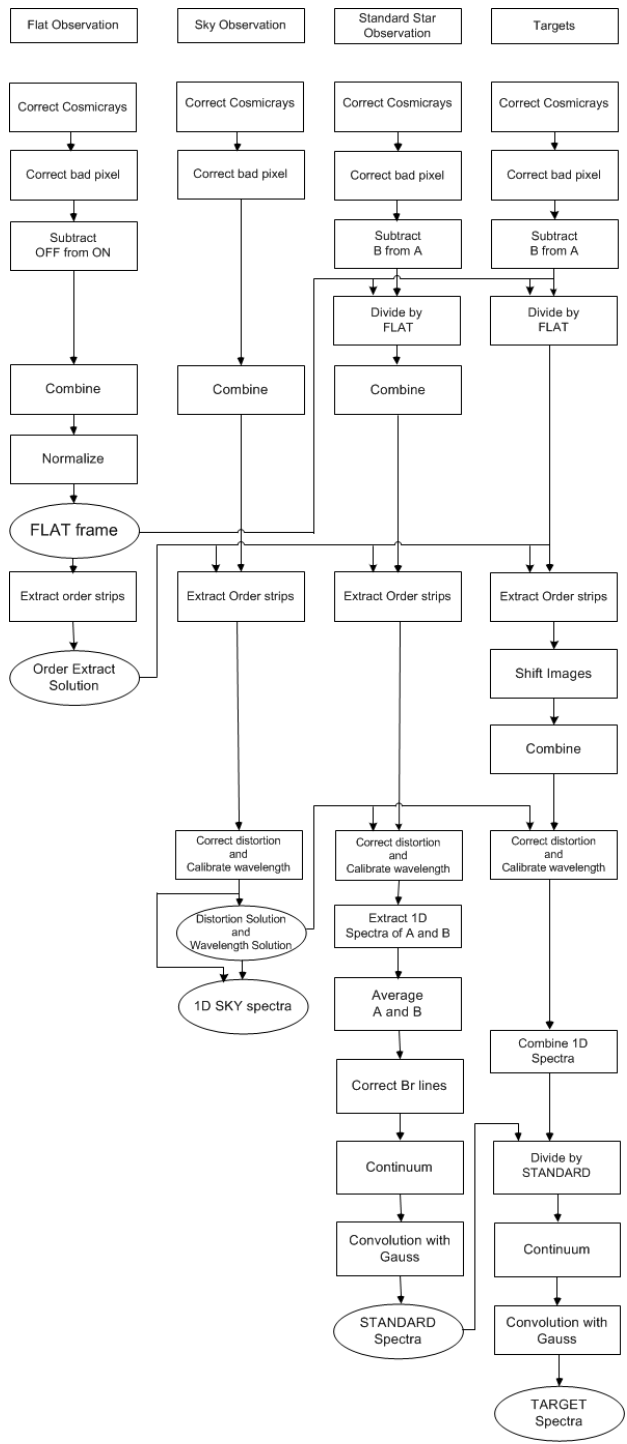}
\caption{Data reduction processes.}
\label{reduce}
\end{figure*}

Data reduction was done by using {\it IRAF} \footnote{{\it IRAF} (Image Reduction and Analysis Facility) is distributed by the National Optical Astronomy Observatories (NOAO).} tasks following the methods described in \citet{Pyo02}. The details of the data reduction for standard stars and template stars can be found in \citet{Le11}. The host galaxy spectra were reduced by using similar procedures as that of the template stars. Fig. \ref{reduce} shows the detailed data reduction processes.

The host galaxy spectra within the radius from 0.24 to 1.89 arcsec are extracted for PG~$0844+039$, and from 0.24 to 2.34 arcsec for PG~$1226+023$ and PG~$1426+015$. We chose the minimum radius to be 0.24 arcsec to ensure that the extracted host galaxy spectra are not affected by emission from QSOs.  We confined the maximum radius to extract the host galaxy spectra within the effective radii \citep{Peng02} of the targets. In the case of PG~$0844+349$, the maximum radius is equal to the effective radius, $R_e$ = 1.89 arcsec. The slit length\footnote{http:$\slash$$\slash$www.naoj.org$\slash$Observing$\slash$Instruments} of Subaru/IRCS, L = 5.17 arcsec, however, is shorter than the diameters of PG~$1226+023$ ($D$ = 12.56 arcsec) and of PG~$1426+015$ ($D$ = 8.22 arcsec). Therefore, the maximum radius to be extracted should be smaller than half of the slit length, L = 2.59 arcsec.

The effects of the residual OH sky-lines could cause one of the problems of our obtained spectra. The emission lines of OH cannot be completely corrected by the sky-background subtraction processes. We masked out the data points which contain noises from the OH sky-lines.

\section{Results and Discussion}

\subsection{The Host Galaxy Spectra}

\begin{sidewaysfigure*}[thp]
\centering
\includegraphics[width=110ex]{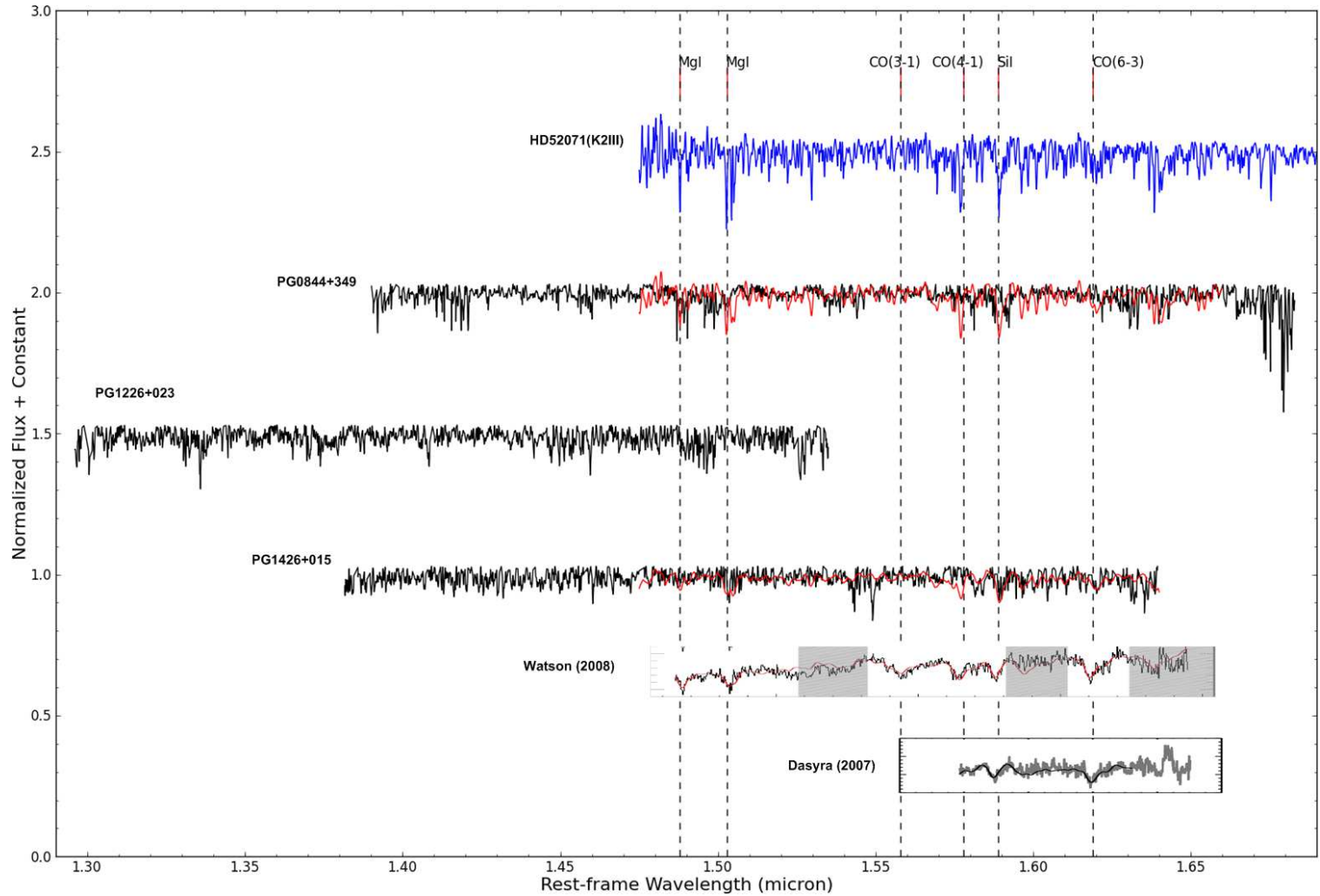}
\caption{Spectra of host galaxy PG~$0844+349$ (shifted with $z = 0.064$), PG~$1226+023$ (shifted with $z = 0.158$), PG~$1426+015$ (shifted with $z = 0.086$), and spectrum of K2 III (HD52071, blue-line). The red-lines show the best fit stellar spectra of K2 III (HD52071). Two spectra at the bottom show the spectra of PG~$1426+015$ from \citet{Watson08} and \citet{Dasyra07}.}
\label{final}
\end{sidewaysfigure*}

Fig. \ref{final} shows the reduced spectra of PG~$0844+349$, PG~$1226+023$, and PG~$1426+015$. In the figure, the spectrum of K2 III (HD$52071$) is also plotted to be compared with the observed host galaxy spectra. In addition, the spectra of PG~$1426+015$ in \citet{Watson08} and \citet{Dasyra07} are shown in the figure for comparisons.

From Fig. \ref{final}, we identify prominent features, e.g., Mg~I (1.488 $\mu$m),  Mg~I (1.503 $\mu$m), Si~I (1.589 $\mu$m), and CO (6-3) (1.619 $\mu$m) in the spectrum of PG~$0844+349$. But the absorption features such as CO (3-0) (1.558 $\mu$m) and CO (4-1) (1.578 $\mu$m) cannot be seen due to the effects of the remaining OH sky-lines.

After the redshift correction, the host galaxy spectrum of PG~$1226+023$ in Fig. \ref{final} has a limited wavelength coverage to be compared with the molecular lines of the stellar template spectrum. Unfortunately, these are not stellar spectra in the literature that overlap with the observed spectrum of PG~$1226+023$ to identify the molecular lines.

In the case of PG~$1426+015$ spectrum, we could detect Mg~I (1.503 $\mu$m), Si~I (1.589 $\mu$m) and CO (6-3) (1.619 $\mu$m) lines comparing the K2 III stellar template spectrum and the host galaxy spectra of \citet{Watson08} and \citet{Dasyra07}. But the CO (3-0) (1.558 $\mu$m) and CO (4-1) (1.578 $\mu$m) lines are hard to confirm because of the effects of the remaining OH sky-lines. The signal-to-noise (S/N) ratios of Mg I and Si I absorption lines are 3. The S/N ratio of CO (6-3) absorption line is 5.

\subsection{Velocity Dispersion}

From the obtained spectra, we identified a few stellar absorption lines of the host galaxy and measured the velocity dispersion of the host galaxy using the direct fitting method of \citet{Barth02}.

We assume that the host galaxy spectrum follows that of K2 III type stars. We define the model of host galaxy spectrum from the equation as

\begin{equation}
M(\lambda )=A + T_{G}\left ( \lambda ,\sigma \right )
\end{equation}

where A is arbitrary constant value from the unknown contribution from the quasar continuum; $T_{G}$ is convolution of the stellar spectrum with the line-of-sight velocity distribution; $\sigma$ is the velocity dispersion; and $\lambda$ is the rest wavelength.
The best-fit convolved stellar spectrum to the host galaxy spectrum is found based on the minimum of chi-square values. The chi-square value is calculated by

\begin{equation}
\chi^{2}=\sum_{\lambda }\left ( \frac{M_{\lambda }-O_{\lambda }}{\epsilon_{\lambda}} \right )^{2}
\end{equation}

where $M_{\lambda }$ is the model spectrum from the equation (1); $O_{\lambda }$ is the observed host galaxy spectrum; and $\epsilon_{\lambda}$ is the error of the host galaxy spectrum which has typical value of 0.037 and is determined by adding the standard deviation of the spectrum to the root-mean-square of each data point.

\begin{figure*}[thp]
\centering
\includegraphics[width=35ex]{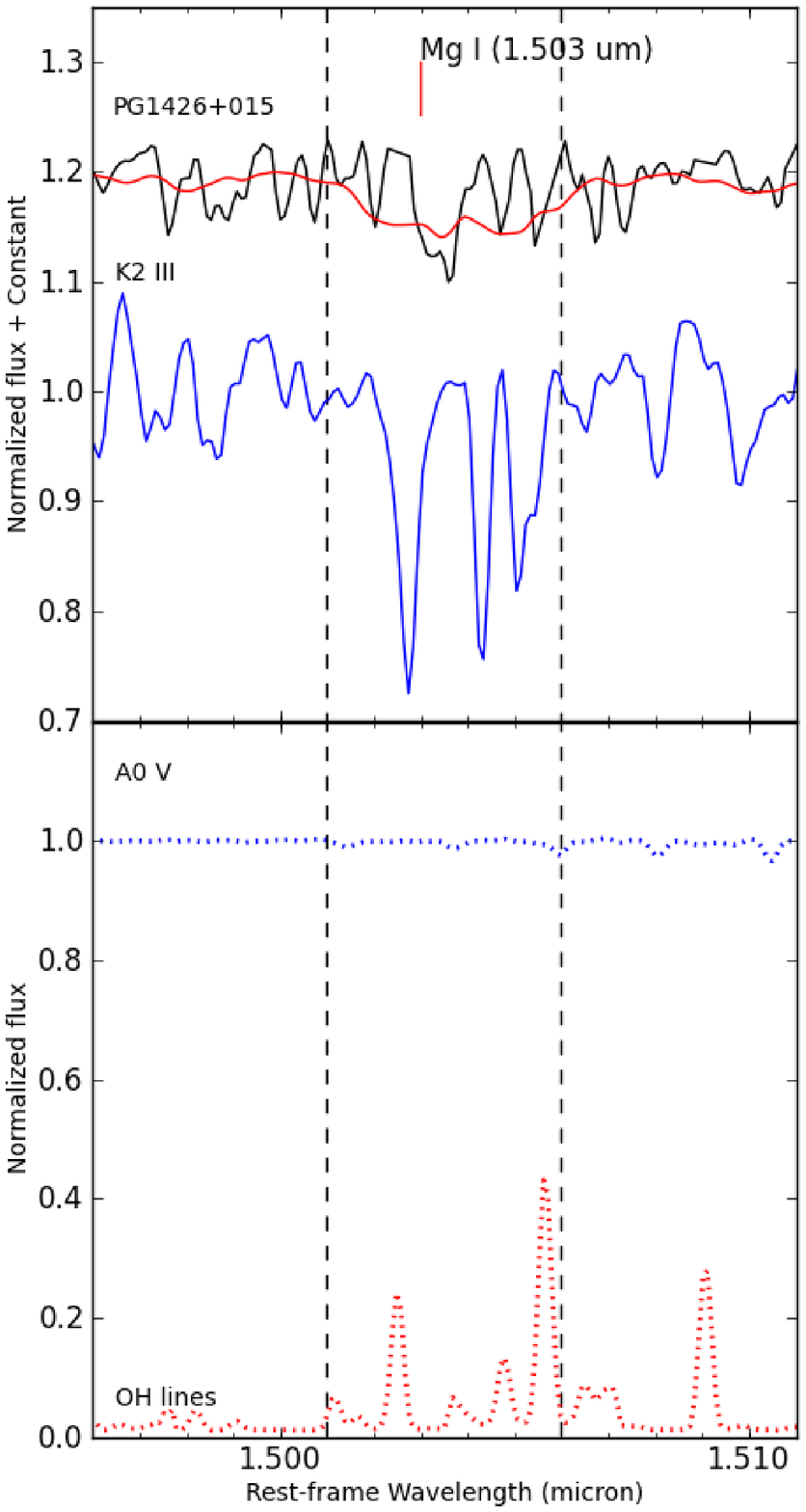}
\caption{Spectra of K2 III (HD52071) (blue-line) and host galaxy PG~$1426+015$ (shifted with $z = 0.086$) (black-line). The best fit of the velocity convolved K2 III spectrum is shown in red-line. The dashed lines show the regions used for measurements of velocity dispersions. The A0 V and OH sky-lines spectra are shown in the lower plot.}
\label{fitting1426}
\end{figure*}

\begin{figure*}[thp]
\centering
\includegraphics[width=35ex]{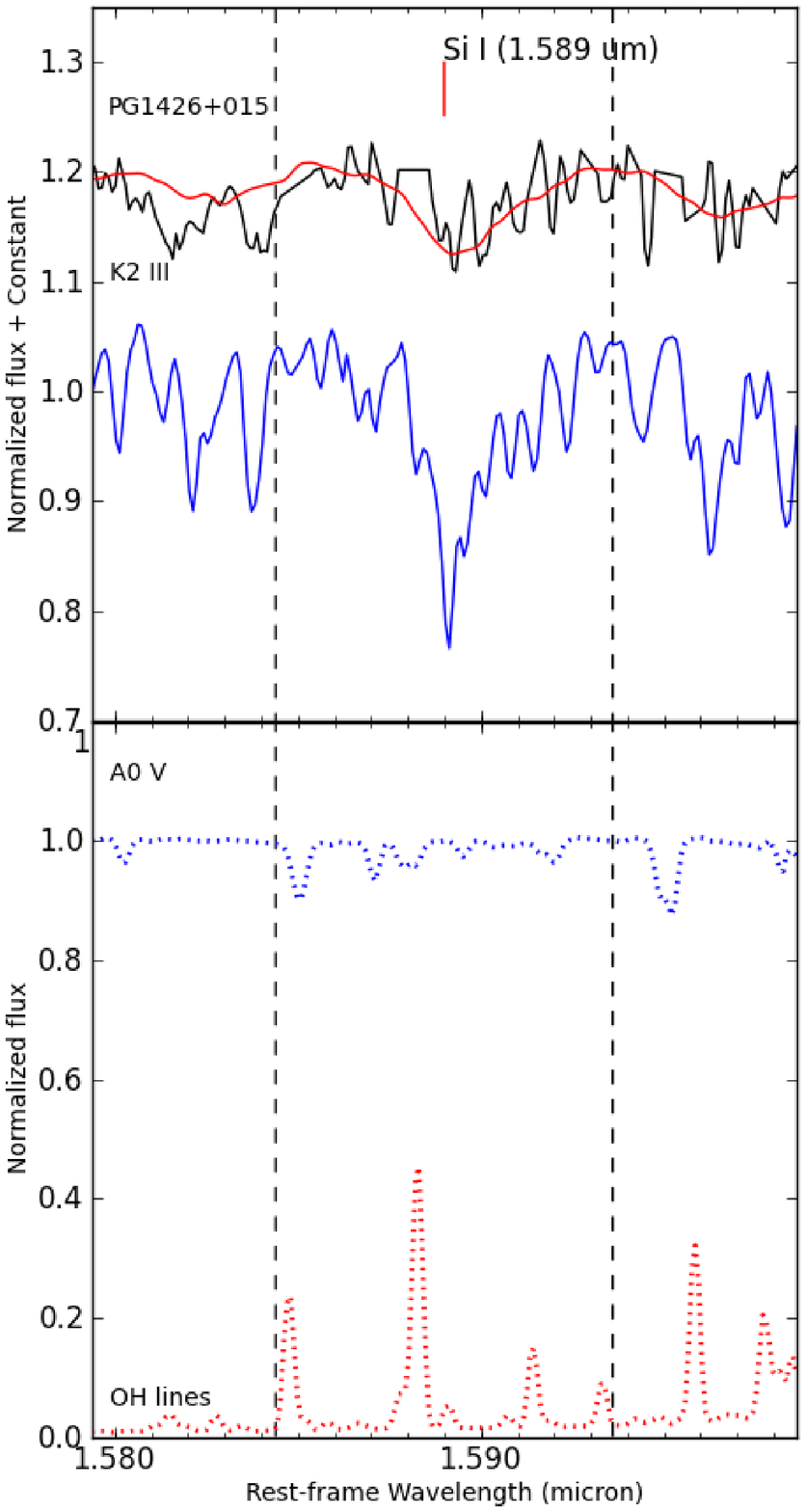}
\caption{Spectra of K2 III (HD52071) (blue-line) and host galaxy PG~$1426+015$ (shifted with $z = 0.086$) (black-line). The best fit of the velocity convolved K2 III spectrum is shown in red-line. The dashed lines show the regions used for measurements of velocity dispersions. The A0 V and OH sky-lines spectra are shown in the lower plot.}
\label{fitting1426si}
\end{figure*}

\begin{figure*}[thp]
\centering
\includegraphics[width=35ex]{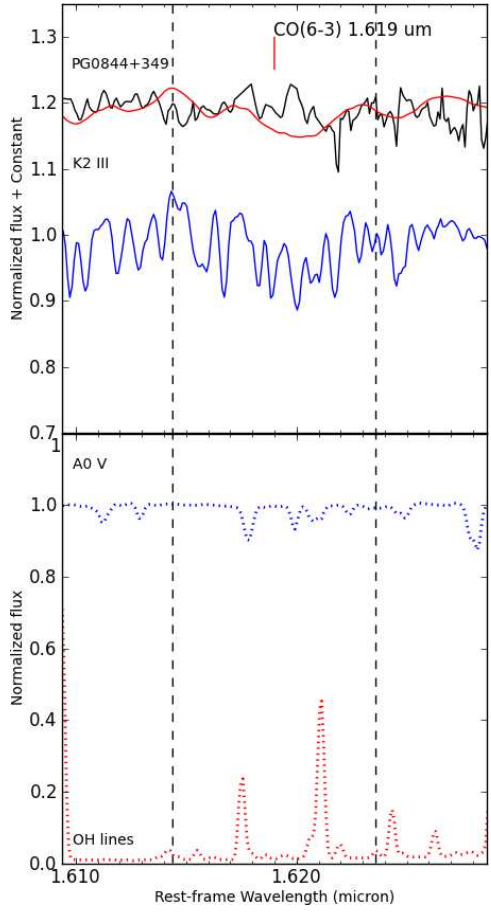}
\caption{Spectra of K2 III (HD52071) (blue-line) and spectra of host galaxy PG~$0844+349$ (shifted with $z = 0.064$). The best fit of the velocity convolved K2 III spectrum is shown in red-line. The dashed lines show the regions used for measurements of velocity dispersions. The A0 V and OH sky-lines spectra are shown in the lower plot.}
\label{fitting0844}
\end{figure*}

From Fig. \ref{final}, due to the redshift correction, we could not measure the velocity dispersion of host galaxy spectrum of PG~$1226+023$. In case of PG$0844+349$, we identified some prominent features since the effects of the remaining OH sky-lines make it hard to calculate the velocity dispersion of the host galaxy. We have corrected the remaining OH sly-lines effects by removing those data points of the host galaxy spectra which are affected by sky-lines. We have calculated the velocity dispersion of the host galaxy PG~$1426+015$ to be $\sigma = 264$ $\pm$ 215 km~s$^{-1}$ from the fitting with K2 III stellar spectrum at Mg~I (1.503 $\mu$m) line (Fig. \ref{fitting1426}) and Si~I (1.589 $\mu$m) line (Fig. \ref{fitting1426si}). The reduced chi-square value is 0.8. In Fig. \ref{fitting0844}, from the measurement of CO (6-3) (1.619 $\mu$m), the velocity dispersion of PG~$0844+349$ is 132 $\pm$ 110 km~s$^{-1}$ with the reduced chi-square of 0.4. Due to the low S/N ratio of the data, the errors of velocity dispersions are very large. But the best-fit sigma values which are calculated from our method are consistent with others. From the measurements of \citet{Watson08} and \citet{Dasyra07}, the velocity dispersions of host galaxy of PG~$1426+015$ are 217 $\pm$ 15 km~s$^{-1}$ and 185 $\pm$ 67 km~s$^{-1}$, respectively, which are similar to our results. The data obtained by ISAAC long-slit spectrometer on the 8m Antu unit of the Very Large Telescope \citep{Dasyra07} has higher S/N ratio compared to our data obtained by IRCS, Subaru telescope.

From the velocity dispersion measurements, we derived the black hole masses using the $M_{BH}$$-$$\sigma$ relation of elliptical galaxies \citep{Kormendy13}. Table \ref{velocity} shows the measured values of velocity dispersion and black hole mass estimates of these quasars. The obtained black hole masses of PG~$0844+349$ and PG~$1426+015$ are $log(M_{BH}/M_{\odot})$  = 7.7 $\pm$ 5.5 and $log(M_{BH}/M_{\odot})$ = 9.0 $\pm$ 7.5, respectively.

Independently, the BH mass can be determined by using the velocity width of a broad emission line and the broad line region size from the reverberation mapping method \citep{Kaspi00}, or the continuum/line luminosity \citep{Kim10}. Assuming a virial factor of $f=5.5$ in \citet{Onken04} and \citet{Woo13}, \citet{Peterson04} find that the BH masses are $log(M_{BH}/M_{\odot})$  = 8.0 $\pm$ 0.2 for PG~$0844+349$ and $log(M_{BH}/M_{\odot})$  = 9.1 $\pm$ 0.2 for PG~$1426+015$.

\section{Conclusions}

We obtained NIR medium resolution host galaxy spectra of nearby quasars, PG~$0844+349$, PG~$1226+023$, and PG~$1426+015$ in H-band, using the IRCS instrument and the AO of the Subaru telescope. The data analysis method of the NIR spectra is presented.

From the spectra, we derived the stellar velocity dispersion of the host galaxy and its relation to the central super-massive BH. From the identified stellar absorption lines, we have obtained the velocity dispersion of PG~$1426+015$ to be 264 $\pm$ 215 km~s$^{-1}$ based on the measurement of Mg~I (1.503 $\mu$m) and Si~I (1.589 $\mu$m). In the case of PG~$0844+349$, we have measured the velocity dispersion of the host galaxy to be 132 $\pm$ 110 km~s$^{-1}$ based on the calculation of CO (6-3) (1.619 $\mu$m).

By using an $M_{BH}$$-$$\sigma$ relation of elliptical galaxies, the BH masses of PG~$0844+349$ and PG~$1426+015$ are estimated to be $log(M_{BH}/M_{\odot})$  = 7.7 $\pm$ 5.5 and $log(M_{BH}/M_{\odot})$ = 9.0 $\pm$ 7.5, respectively. These values are consistent with the BH masses from the quasar broad emission lines. \\


This work was supported by the National Research Foundation of Korea (NRF) grant, No. 2008-0060544, funded by the Korea government (MSIP). We appreciate Tae-Soo Pyo and Hiroshi Terada for observations and data reductions, and Elaine S. Pak for proofreading this manuscript. LCH acknowledges support from the Chinese Academy of Science through grant No. XDB09030102 (Emergence of Cosmological Structures) from the Strategic Priority Research Program.
This work is based on the data collected at Subaru Telescope, which is operated by the National Astronomical Observatory of Japan.

\clearpage

%



\clearpage

\begin{table*}[thp]
\caption{Observation log}
\centering
\doublerulesep1.0pt
\renewcommand\arraystretch{1.5}
\resizebox{16cm}{!}{
    \begin{tabular}{ccccccc}
    \hline
    Date      &   Quasars   &  z  &    R  &  EchelleSetting  &  Slit Width      &    Total Exposure \\
    (UT)      &             &     &       &                   &   (arcsec)       &    (sec)  \\
    \hline
    2003 Feb 11    & PG~$0844+349$   &  0.064       &     10000   &  $H+$  &  0.3     &  $37 \times 180$  \\ [0.5ex]
    2004 April 3   & PG~$0844+349$   &  0.064        &     5000    &  $H-$  &  0.6     & $10 \times 300$  \\
    2004 April 3   & PG~$1226+023$   &  0.158       &   5000    &  $H-$  &  0.6     &  $8 \times 300$   \\
    2004 April 3   &  PG~$1226+023$  & 0.158       &   5000    &  $H+$  &  0.6      &  $8 \times 300$  \\
    2004 April 3  & PG~$1426+015$   &  0.086       &   5000    &  $H-$  &   0.6     &  $5 \times 300$  \\
    2004 April 3  & PG~$1426+015$ &   0.086      &   5000    &  $H+$  &  0.6      &  $5 \times 300$  \\[0.5ex]
    2004 April 4   & PG~$0844+349$   &  0.064       &    5000    &  $H+$  &  0.6      &     $8 \times 300$  \\
    2004 April 4        & PG~$1426+015$   &  0.086       &   5000    &  $H-$  & 0.6       &  $12 \times 300$   \\
    2004 April 4        & PG~$1426+015$ &   0.086       &   5000    &  $H+$  &  0.6      &    $15 \times 300$ \\[0.5ex]
    \hline
    \end{tabular}
    }
\label{observation}
\end{table*}

\clearpage

\begin{table*}[thp]
\caption{Velocity dispersion measurements and black hole masses}
\doublerulesep1.0pt
\renewcommand\arraystretch{1.5}
\resizebox{12cm}{!}{
    \begin{tabular}{ccccc}
    \hline
    Quasars   &  $\sigma$$^{a}$      &  $\sigma$$^{b}$              &     $log(M_{BH}/M_{\odot})$$^{c}$            &      $log(M_{BH}/M_{\odot})$$^{d}$  \\
              &   (km~s$^{-1}$)     &        (km~s$^{-1}$)            &              &      \\
    \hline
    PG~$0844+349$   &  132 $\pm$ 110$^{e}$      &              &     7.7 $\pm$ 5.5  &    8.0 $\pm$ 0.2  \\
    PG~$1426+015$   &  264 $\pm$ 215$^{f}$     &    217 $\pm$ 15    &  9.0 $\pm$ 7.5  &  9.1 $\pm$ 0.2   \\ [0.5ex]
    \hline

    \end{tabular}

    }

    \scriptsize{$^{\rm a}$ This work.} \\
    \scriptsize{$^{\rm b}$ Velocity dispersion values from \citet{Watson08}.} \\
    \scriptsize{$^{\rm c}$ This work. Black hole mass values are derived by using the formula (7) of \citet{Kormendy13}.} \\
    \scriptsize{$^{\rm d}$ Black hole mass values from \citet{Peterson04}.} \\
    \scriptsize{$^{\rm e}$ Velocity dispersion value based on the measurement of CO (6-3) (1.619 $\mu$m) line.} \\
    \scriptsize{$^{\rm f}$ Velocity dispersion value based on the measurements of Mg~I (1.508 $\mu$m) and Si~I (1.589 $\mu$m) lines.} \\

\label{velocity}
\end{table*}

\clearpage

%

\clearpage

\clearpage

\clearpage

\clearpage

\end{document}